\documentclass[twocolumn]{webofc}
\usepackage[varg]{txfonts}   % Web of Conferences font
\usepackage{epsfig,graphicx,float,color}
\usepackage{amssymb}
\usepackage{amsmath}
\usepackage[normalem]{ulem}
\begin{document}

\title{Sr isotopes: the interplay between shape coexistence and quantum phase transitions}

\author{\firstname{E.} \lastname{Maya-Barbecho}\inst{1}\fnsep\thanks{\email{esperanza.maya@alu.uhu.es}} \and
        \firstname{J.E.} \lastname{Garc\'{\i}a-Ramos}\inst{1,2}\fnsep\thanks{\email{enrique.ramos@dfaie.uhu.es}}}

\institute{Departamento de  Ciencias Integradas y Centro de Estudios Avanzados en F\'{\i}sica, Matem\'atica y Computaci\'on, Universidad de Huelva, 21071 Huelva, Spain
\and Instituto Carlos I de F\'{\i}sica Te\'orica y Computacional,  Universidad de Granada, Fuentenueva s/n, 18071 Granada, Spain }

\abstract{In this contribution we study the even-even Sr isotopes considering the influence of particle-hole intruder states. We find that the rapid onset of deformation at A $=98$ and onwards can be explained by the crossing of regular and intruder states. We compare the systematics of Sr with the ones of Zr, Pt, and Hg nuclei.} 

\maketitle

\section{Introduction}
\label{intro}
The understanding of the appearance of nuclear deformation is still of major interest in nuclear physics. The collective phenomena in the atomic nucleus is the origin of the onset of deformation, but it is also the responsible of the vibrational character of nuclei. The range of phenomena between both limits is modulated by a fine balance between the pairing, which tends to produce spherical shapes, and the quadrupole force, which makes the nucleus deformed. In others words, there is a competition between the stabilizing effect of the shell closure energy gap and the residual interaction between like and unlike nucleons.

Nuclei near to a shell closure tends to be spherical and excitations are relatively costly, specially the particle-hole ones, but, surprisingly, enough of the energy of such excitations can be considerably reduced due to the increase of the residual interaction because of the presence of a larger number of effective valence nucleons. This effect is particularly important when protons are near to a shell closure, while neutrons are at the middle of a shell, or viceversa. In this case, a particle-hole excitation will enhance the proton-neutron quadrupole interaction and it will considerable reduce the energy of such an excitation, being possible to appear at very low energy, even becoming the ground state of the system. Moreover, these configurations are expected to be more deformed than the normal ones. We will refer to them as intruder and regular configurations, respectively, and to the interplay between them as configuration mixing \cite{heyde11}. %There exist many examples where intruders appear at very low energy presenting a much more deformed character than the regular states \cite{Poves16}.

Along this contribution we will study the connection between shape coexistence and the so called Quantum Phase Transition (QPT) \cite{Cejn09}. We will focus on the case of Sr, but a detailed comparison with Zr, Pt, and Hg nuclei is included.
%The very rapid onset of deformation can be also understood in terms of the so called Quantum Phase Transition (QPT) \cite{Cejn09}, which supposes the existence of an abrupt change in the structure of the nucleus (deformation, radius or binding energy, among other properties) under a small change of a control parameter (the neutron number, for instance).

\section{Shape coexistence and quantum phase transitions: faces of a single phenomenon?}
\label{sec-sc-qpt}
Shape coexistence in nuclear physics was first proposed by Morinaga in the 1950's % \cite{morinaga56} 
and nowadays becomes a concept that appears throughout the entire nuclear landscape, especially in those nuclei at or near shell or sub-shell closures \cite{heyde11}. Shape coexistence presents certain key experimental features: an U shape pattern in the energy systematics of a set of states, the lowering of certain excited $0^+$ states (intruder states), a rapid change in the value of the spectroscopic quadrupole moments, and the existence of strong E0 transitions. All of them show an almost symmetric trend with respect to the corresponding mid-shell and they are enhanced at this place. 

Shape coexistence can be understood in terms of two well-known theoretical approaches: the spherical shell model and the mean field. According to the shell model approach, % the nucleus is described as an inert core and a set of valence nucleons that occupy certain orbits and interact amongst them through a residual two-body interaction. 
the intruder configurations correspond to the promotion of pairs of nucleons across the shell or subshell gap closures. The other major approach is the mean-field theory where, roughly speaking, one obtains an energy surface depending on certain deformation parameters. In this landscape, the ground band will correspond to the deepest minimum (regular states), but the other local minima can be interpreted as intruder configurations, each of them owning to different deformations.

The concept of QPT \cite{Cejn09} implies the sudden change of the ground state structure of the system as a function of a control parameter. Such a control parameter can be in the case of the atomic nucleus, e.g., the neutron number. Therefore, a QPT can appear in an isotopic chain where the ground state deformation varies in an abrupt way when passing from isotope to isotope. Of course, the latter should be considered as an approximation because a control parameter should be, strictly speaking, continuous.
A QPT is characterized by the discontinuity of some derivative of the system energy. Assuming that the two-neutron separation energy (S$_{2n}$) is somehow proportional to the derivative of the energy with respect to the neutron number, a first-order QPT involves a discontinuity in S$_{2n}$, while a second-order one a discontinuity in the derivative of S$_{2n}$. The even-even Sr isotope chain, as will be shown along the text, corresponds to the case of a first-order QPT.

An ideal framework to deal with QPTs is the interacting boson model (IBM) \cite{iach87} which is a paradigmatic example of a nuclear algebraic model. In that framework, a QPT can be understood considering a Hamiltonian that is a combination of two given symmetries, combined through a control parameter, $H=x\ H_{\mbox{sym1}}+(1-x)\ H_{\mbox{sym2}}$ that allows to go from one to the other limit. The presence of a QPT is revealed by the existence of a critical value of the control parameter, x$_c$, for which the underlying structure of the system (phase) passes from {\it symmetry 1} to {\it symmetry 2}.

The IBM can be also used to deal with shape coexistence, allowing to define a framework where different particle-hole excitations are present in the spectrum. The formalism is known as IBM with configuration mixing, IBM-CM in short \cite{duval81,duval82}. 

Is there any connection between shape coexistence and QPT? First, one should note several similarities, namely, both phenomena involve a rapid change in the structure of certain set of states, either ground or excited states and, in both cases, at the mean-field level, several minima coexist. 
In the case of shape coexistence, the regular and the intruder configurations can be thought as having independent energy surfaces that interact among them, especially when their minima are close in energy, while under a QPT a single energy surface exists where two or more minima are present. The evolution of the relative position of the minima as a function of the neutron number is at the origin of the change of the nuclear deformation and it can be understood in terms of either shape coexistence or a QPT. 
%The situation becomes even more challenging when in the case of shape coexistence the regular and the intruder energy surfaces also evolve \cite{Gavr19}. 
On the other hand, from a quantum point of view clear differences should exist between both approaches. The most obvious one is the existence of a larger Hilbert space when two configurations are present, because of the extra nucleons that occupy the single particle levels. This extra set of levels presents a different energy systematics than the regular one because for them the effective nuclear interaction is different. Moreover, the intruder configuration is expected to be found at low energy when the number of valence nucleons is large, i.e. around mid shell, therefore a parabolic-like energy systematics for the intruder states appears frequently. However, the situation can be blurred due to the interaction between intruder and regular configurations or because their crossing \cite{Garc11}. When a QPT exist without the presence of multiple particle-hole excitations, a larger density of states can be also observed at low energy as for configuration mixing, but it never happens around mid-shell \cite{Garc20}. The remaining open question is whether a QPT can be induced by shape coexistence, which we will try to answer along the rest of the contribution.

\section{Sr and Zr isotopes versus Pt and Hg nuclei}
\label{sec-Sr}
The regions around the sub-shell closure Z $=40$ are, together with the area around Z $=82$, two of the best examples of shape coexistence in nuclei. In Hg and Pt nuclei, two types of configurations show up with a clear presence of low lying $0^+$ states, combined with a parabolic-like shape in the spectra of Hg, whereas this behavior is not so obvious in Pt. On the other hand, Zr and Sr are known to exhibit the faster change in deformation in an isotope chain at N $=60$, with two particle-hole configuration coexisting and crossing at this point.
%, being $^{96}$Sr ($^{98}$Zr) almost spherical and $^{98}$Sr ($^{100}$Zr) a very well deformed nucleus. Concerning energy systematics, both Pt and Hg isotopes (but also Pb or Po) (see for instance \cite{Garc09,Garc11,Garc14b,Garc15}) present a clear symmetric behavior with respect to the mid-shell, N $=104$, however, in the case of Sr and Zr this symmetry cannot be observe due to the lack of experimental information for very neutron-rich nuclei. Besides, a sudden change of the nuclear structure is observed at N $=60$, clearly visible because of the dropping of the E$(2_1^+)$ energy or the rising of the E$(4_1^+)$ /E$(2_1^+)$ ratio (see for instance \cite{Garc19,Garc20,Maya22} and Fig.~\ref{fig-ratio}). 
\begin{figure}[h]
\centering
\includegraphics[width=0.44\textwidth]{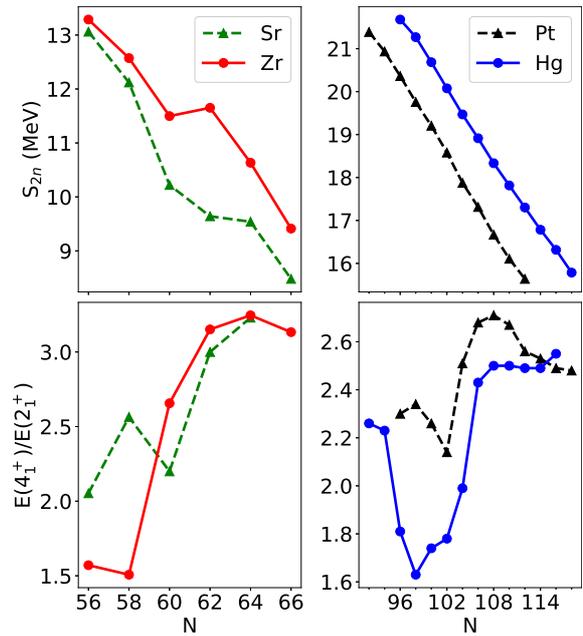}
\caption{S$_{2n}$ and $E(4_1^+)/E(2_1^+)$ as a function of the neutron number, $N$, for Sr, Zr, Pt and Hg isotopes.}
\label{fig-ratio}
\end{figure}

From the point of view of a QPT, there are two key observables that can be used as indicators of its presence, namely, the two-neutron separation energy, S$_{2n}$, that will experience a discontinuity at a first-order transition point and the ratio $E(4_1^+)/E(2_1^+)$ which provides a hint for the appearance of deformation or, in a qualitative way, it is connected with the order parameter of the QPT. In Fig.~\ref{fig-ratio}, we present the systematics of S$_{2n}$ and $E(4_1^+)/E(2_1^+)$, for Sr and Zr, on the one hand and, for Pt and Hg, on the other. Regarding the value of S$_{2n}$,  one notices a clear change of slope in Sr and Zr, but a fully linear behavior in Hg and Pt. On the other hand, $E(4_1^+)/E(2_1^+)$ shows for Sr and Zr the typical rapid change that is expected in a QPT, passing from a value close to $2$ (even $1.5$ for $^{96-98}$Zr) to another close to $3.3$. However, in Pt and Hg the interpretation of the systematics is not so evident and the physics involved in these cases seems to be quite different from the one of Sr and Zr. 
\begin{figure}
\centering
\sidecaption
\includegraphics[width=0.44\textwidth]{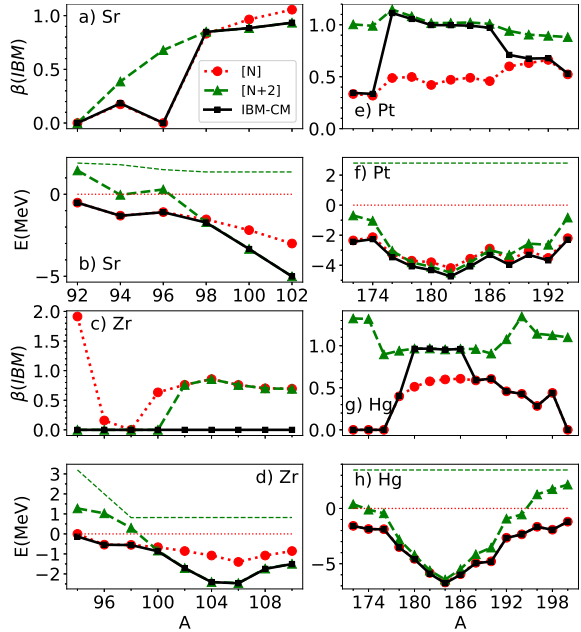}
\caption{a), c), e), and g) panels: IBM value of $\beta$ for the unperturbed regular, $[N]$, and intruder configurations, $[N+2]$, and full calculation, IBM-CM. b), d), f), and h) panels: ground state energy for the unperturbed regular and the intruder configurations and for the full calculation. The green thin dashed line stands for the energy needed to create a particle-hole excitation and the red thin dotted one stands for the reference level.}
\label{fig-beta-energy}
\end{figure}

To have a more clear view of the analogies and differences between both situations, we present in Fig.~\ref{fig-beta-energy} some relevant IBM calculations based on \cite{Maya22} for Sr, \cite{Garc20} for Zr, \cite{Garc09} for Pt, and \cite{Garc14b} for Hg. In particular, the results correspond to the value of the IBM deformation parameter $\beta$ for the regular ([N]), the intruder configuration ([N+2]) and the complete system (IBM-CM) (panels a), c), e), and g)). Moreover, the unperturbed energy for the lowest regular, intruder, and ground state (IBM-CM) are depicted in panels b), d), f), and h). Concerning Sr and Zr, the distinctive feature is the crossing of the regular and the intruder configurations at A $=98$ and $100$, respectively, corresponding to N $=60$, inducing a change in the slope of the ground state energy and also a rapid onset of deformation, which has a dramatic effect on the trend of the radii \cite{Garc19,Maya22}. In the case of Pt and Hg, both configurations remain quite close, without crossing in the case of Hg, and crossing before and after the mid-shell (N $=104$) in the case of Pt. The ground state energy is hardly modified, but the deformation parameter $\beta$ passed from the less deformed configuration (the regular one) to the most deformed one (the intruder one) around mid-shell. The crossing or the close approaching of regular and intruder configurations in Pt and Hg, respectively, have also a noticeable influence in the radii of these nuclei, as shown in \cite{Garc11,Garc14b}. In summary, in Sr, Zr and Pt nuclei, a intruder state becomes the ground state when approaching the mid-shell, while it never happens in Hg. Note that in Pt, due to the strong mixing between both configurations \cite{Garc09}, no dramatic effect are observed in the spectra, although the energy crossing exists. 

%In Sr, Zr and Pt nuclei, a intruder state becomes the ground state around the mid-shell, while it never happens for Hg. That means that the structure of the ground state changes abruptly in Sr, Zr and Pt, but not in Hg. In Hg a change in the structure is observed in the $2^+$ states. \cite{Garc14b}. Note that the more realistic calculation involves a certain degree of mixing between the regular and the intruder states, which is rather small in Sr \cite{Maya22}, Zr \cite{Garc19} and in Hg \cite{Garc14b}, but much larger in Pt \cite{Garc09}, which means that the observed behavior in Pt is smoothed out in the more realistic calculation, but remains essentially unchanged in Sr, Zr, and Hg.
\begin{figure}
\centering
\sidecaption
\includegraphics[width=0.44\textwidth]{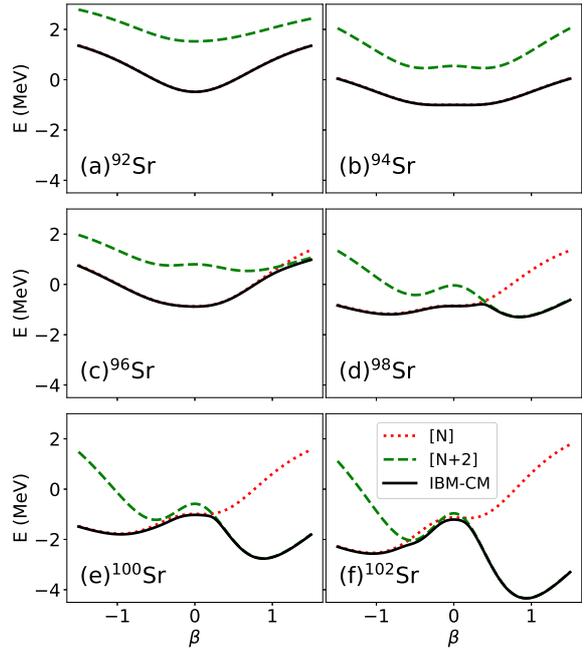}
\caption{Axial IBM-CM mean-field energy surfaces for $^{92-102}$Sr isotopes. Red dotted  line stands for the unperturbed regular configuration, [N], green dashed  one for the unperturbed intruder configuration, [N+2], while the black continuous line for the full IBM-CM calculation.}
\label{fig-energy-surf} 
\end{figure}

The above features have direct consequences in the systematics of S$_{2n}$ and $E(4_1^+)/E(2_1^+)$. In the case of Sr and Zr, what it is observed in Fig.~\ref{fig-ratio} can be easily understood in terms of the crossing of configurations. Indeed, the change in the nature of the ground state produces a sudden change in the slope of the energy systematics and therefore a discontinuity in the value of S$_{2n}$. On the other hand, because the crossing is so fast, not only the ground state presents an intruder character, but also the rest of low-lying members of the yrast band involved in the energy ratio $E(4_1^+)/E(2_1^+)$, which makes that for $N<60$ all the states present a vibrational character, while for $N\ge 60$  a rotational one appears. On the other hand, in the case of Hg, because the ground state presents all the way a regular character and, moreover, there is no interaction between intruder and regular sector, it  can be understood the linear tendency of the S$_{2n}$ even at the mid-shell. Besides, the crossing between $2_1^+$ and $4_1^+$ regular and intruder states can explain the drop of the ratio $E(4_1^+)/E(2_1^+)$ around the mid-shell. Finally, to explain the
systematics of Pt in Fig.\ref{fig-ratio} is a challenge. First, the crossing of regular and intruder $0^+$ energies would suggest a discontinuity in S$_{2n}$ as the observed one in Sr and Zr. However, experimental values show a fully linear systematics. The reason for such a behavior is the strong mixing between the regular and the intruder sectors (see \cite{Garc09}). Note that according to \cite{Garc14a} and figure Fig.~\ref{fig-beta-energy}, the deformation changes as it should be in a QPT, however the strong mixing somehow hides its effect in both the S$_{2n}$ and $E(4_1^+)/E(2_1^+)$ observables.  

A different way of seen the evolution of the nuclear deformation is to study the mean-field energy surfaces in a chain of isotopes, which can be calculated also using the IBM-CM formalism \cite{Frank02}. This is depicted in Fig.~\ref{fig-energy-surf} for Sr isotopes, where the axial energy surfaces for the regular and the intruder configurations, as well as the one for the complete system are presented. Concerning the regular configuration (red dotted line), it presents a smooth evolution being spherical for A $=92-96$, becoming very flat and finally axially oblate deformed for A $= 98-102$. Regarding the intruder configuration (green dashed line), it starts being also spherical for A $=92$ and $94$, but rapidly becomes axially prolate deformed from A $=96$ and onwards. Finally, the IBM-CM full calculation (black full line), which essentially follows the lowest of the intruder and regular curves, starts also with a spherical minimum, A $=92-96$ and also becomes an axially prolate deformed one for A $=98-102$. A very relevant feature is that both configurations evolve separately from a spherical to a deformed shape, more rapidly in the case of the intruder configuration. Therefore, most probably in both configurations separately, regular and intruder, a QPT is present. Moreover, the fact that both configurations cross at A $=98$ makes much more abrupt the change of deformation, also inducing a QPT in the whole system. The QPT present in the separated regular and the intruder configurations is called type-I QPT \cite{Gavr19} and involves the existence of a single configuration energy surface where several minima are present. On the other hand, for the whole energy surface a type-II QPT \cite{Gavr19} appears because two energy surfaces that cross are present. A type-II QPT is expected to induce much more abrupt changes than the type-I and it will own the typical phenomenology of a first order QPT. Note that the type-I QPT is also able to reproduce the S$_{2n}$ trend of Zr isotopes, as was shown in \cite{Garc05}, although the type-II QPT provides a more consistent framework \cite{Garc20} to explain the rapid onset of deformation. It is worthy to mention that this is not the only mechanism to induce a discontinuity in the S$_{2n}$, as happens, for instance, in the rare-earth region at $N\approx 90$.

\section{Conclusions}
\label{sec-conclu}
In this contribution, we have studied the onset of deformation in Sr on the light of the coexistence of several particle-hole configurations. We have compared with Zr isotopes that is another example of $Z\approx 40$ nucleus and that presents a common trend regarding many observables and with Pt and Hg nuclei that are paradigmatic examples of nuclei where shape coexistence plays a major role in $Z\approx 82$. With the latter case notable differences exist.

The origin of the onset of deformation in $Z\approx 40$ region can be understood in terms of the residual proton-neutron interactions resulting in major modifications in the occupation of the 1g$_{9/2}$ proton and 1g$_{7/2}$ neutron orbitals. This approach suggested in Federman and Pittel \cite{Fede77,Fede79b} emphasizes the importance of the simultaneous occupation of neutrons and protons spin-orbit partners. In this work, we have tried to define an effective approach for the Federman-Pittel mechanism using the IBM-CM.

In the framework of the IBM, the rapid changes observed in transitional regions, as is the case of Sr, can be understood in terms of a Quantum Phase Transition (QPT), however, the use of configuration mixing, IBM-CM, explains in a more consistent way the very abrupt changes in terms of the crossing of two configurations, calling to this situation type-II QPT. Note that the crossing does not always creates a QPT, as happens in the case of Pt, where the strong interaction between the involved particle-hole configurations hinder the observation of abrupt changes.  

Sr, together with Zr, isotopes are very clear examples of QPTs that can be explained in terms of the crossing of two configurations, in other words, examples of Type-II QPTs. 

\section{Acknowledgements}

 This work was partially supported by grant PID2019-104002GB-C21 funded by MCIN/AEI/10.13039/50110001103 and ``ERDF A way of making Europe''. Resources supporting this work were provided by the CEAFMC and Universidad de Huelva High Performance Computer (HPC@UHU) funded by ERDF/MI\-NE\-CO project UNHU-15CE-2848.

\bibliography{references-IBM-CM,references-QPT}

\begin{thebibliography}{17}

\bibitem{heyde11}
K.~Heyde, J.L. Wood, Rev. Mod. Phys. \textbf{83}, 1467 (2011)

\bibitem{Cejn09}
P.~Cejnar, J.~Jolie, Progress in Particle and Nuclear Physics \textbf{62}, 210
  (2009)

\bibitem{iach87}
F.~Iachello, A.~Arima, \emph{{The interacting boson model}} (Cambridge
  University Press, Cambridge, 1987)

\bibitem{duval81}
P.D. Duval, B.R. Barrett, Phys. Lett. B \textbf{100}, 223  (1981)

\bibitem{duval82}
P.D. Duval, B.R. Barrett, Nucl. Phys. A \textbf{376}, 213  (1982)

\bibitem{Garc11}
J.E. Garc\'{\i}a-Ramos, V.~Hellemans, K.~Heyde, Phys. Rev. C \textbf{84},
  014331 (2011)

\bibitem{Garc20}
J.E. Garc\'{\i}a-Ramos, K.~Heyde, Phys. Rev. C \textbf{102}, 054333 (2020)

\bibitem{Maya22}
E.~Maya-Barbecho, J.E. Garc\'{\i}a-Ramos, Phys. Rev. C \textbf{105}, 034341
  (2022)

\bibitem{Garc09}
J.~Garc\'{\i}a-Ramos, K.~Heyde, Nucl. Phys. A \textbf{825}, 39  (2009)

\bibitem{Garc14b}
J.E. Garc\'{\i}a-Ramos, K.~Heyde, Phys. Rev. C \textbf{89}, 014306 (2014)

\bibitem{Garc19}
J.E. Garc\'{\i}a-Ramos, K.~Heyde, Phys. Rev. C \textbf{100}, 044315 (2019)

\bibitem{Garc14a}
J.E. Garc\'{\i}a-Ramos, K.~Heyde, L.M. Robledo, R.~Rodr\'{\i}guez-Guzm\'an,
  Phys. Rev. C \textbf{89}, 034313 (2014)

\bibitem{Frank02}
A.~Frank, O.~Casta\~nos, P.V. Isacker, E.~Padilla, AIP Conf. Proc.
  \textbf{638}, 23 (2002)

\bibitem{Gavr19}
N.~Gavrielov, A.~Leviatan, F.~Iachello, Phys. Rev. C \textbf{99}, 064324 (2019)

\bibitem{Garc05}
J.E. Garc\'{\i}a-Ramos, K.~Heyde, R.~Fossion, V.~Hellemans, S.~De~Baerdemacker,
  Eur. Phys. J. A \textbf{26}, 221 (2005)

\bibitem{Fede77}
P.~Federman, S.~Pittel, Phys. Lett. B \textbf{69}, 385  (1977)

\bibitem{Fede79b}
P.~Federman, S.~Pittel, Phys. Rev. C \textbf{20}, 820 (1979)

\end{thebibliography}
\end{document}